# Magnetic properties and energy absorption of $CoFe_2O_4$ nanoparticles for magnetic hyperthermia

**T.E. Torres**[1,2,3], **A.G. Roca**[1,4], **M.P. Morales**[5], **A. Ibarra**[1], **C. Marquina**[2,3], **M.R. Ibarra**[1,2,3] and **G.F. Goya**[1,2]

[1] Aragon Institute of Nanoscience (INA), University of Zaragoza, Zaragoza Spain.
[2] Condensed Matter Department, Sciences Faculty, University of Zaragoza, Spain.
[3] Instituto de Ciencia de Materiales de Aragón (ICMA), CSIC - Universidad de Zaragoza, Zaragoza, Spain.
[4] Networking Biomedical Research Center (CIBER-BBN), Zaragoza, Spain.
[5] Instituto de Ciencia de Materiales de Madrid, CSIC, Madrid, Spain.

**Abstract.** We have studied the magnetic and power absorption properties of three samples of $CoFe_2O_4$ nanoparticles with sizes from 5 to 12 nm prepared by thermal decomposition of Fe (acac)$_3$ and Co(acac)$_2$ at high temperatures. The blocking temperatures $T_B$ estimated from magnetization M(T) curves spanned the range $180 \leq T_B \leq 320$ K, reflecting the large magnetocrystalline anisotropy of these nanoparticles. Accordingly, high coercive fields $H_C \approx 1.4 - 1.7$ T were observed at low temperatures. Specific Power Absorption (SPA) experiments carried out in ac magnetic fields indicated that, besides particle volume, the effective magnetic anisotropy is a key parameter determining the absorption efficiency. SPA values as high as 98 W/g were obtained for nanoparticles with average size of $\approx 12$ nm.

## 1. Introduction.

The capability of magnetic nanoparticles (MNPs) to act as effective heating agents for Magnetic Hyperthermia (MHT) was demonstrated many years ago [1]. Considerable efforts have been made in the synthesis and characterisation of MNPs to assess their capacities as heat generating agents and to establish the mechanisms governing heat generation at the nanoscale. Several studies have shown a link between the MNPs energy absorption in ac magnetic fields and their size [2]. We present a study of $CoFe_2O_4$ single-domain MNPs, having sizes from 5 to 12 nm, and relate their structural and magnetic properties with the power absorption efficiency.

## 2. Experimental Procedure

$CoFe_2O_4$ nanoparticles of different sizes were prepared by high temperature decomposition of iron and cobalt organic precursors as described elsewhere [3,4]. The samples were synthesized using iron and cobalt acetylacetonate and different solvents (phenyl ether and 1-octadecene) which led to different synthesis temperatures. To control the final particle size, different precursor/surfactant molar ratios were used. Table summarizes the different synthesis conditions. The resulting nanoparticles were washed several times with ethanol after magnetically-assisted precipitation, and the final product was re-dispersed in hexane.

**Table 1:** *Molar precursor/surfactant ratio $R_{mol}$, reflux temperature $T_{ref}$ and growth time $t_G$ at $T_{ref}$, used for the synthesis of the $CoFe_2O_4$ nanoparticles.*

| Sample | solvent | $R_{mol}$ [precursor]/[surfactant] | $T_{ref}$ (ºC) | $d_{TEM}$ (nm) | $\sigma$ |
|---|---|---|---|---|---|
| M01 | Diphenyl-ether | 1:3 | 260 | 5.7(3) | 0.26(3) |
| M03 | Diphenyl-ether | 1:10 | 260 | 7.0(4) | 0.25(4) |
| S01 | 1-Octadecene | 1:3 | 360 | 12.7(4) | 0.16(2) |

Particle size and shape were studied by Transmission Electronic Microscopy (TEM) using a thermoionic 200 kV Tecnai T20 microscope. The mean particle size ($d_{TEM}$) and statistical size distribution were evaluated by measuring the largest internal dimension of at least 100 particles. X-ray diffraction of powders were made between 10º < 2θ < 70º using Cu-Kα radiation. Iron and cobalt concentrations were determined by Atomic Emission Spectroscopy-Inductively Coupled Plasma (AES-ICP).

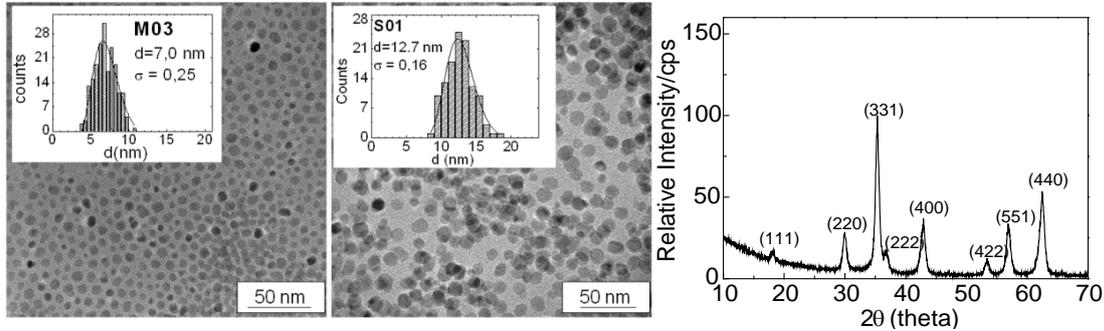

**Figure 1**. (a) TEM images of CoFe2O4 nanoparticles obtained from decomposition of Fe(III) and Co(II) acetylacetonate in Fenhyl ether (sample M03, left) and in 1-Octadecene (sample S01, right); **Insets:** log-normal size-distributions. (b) X-ray pattern of sample S01.

Magnetization and ac susceptibility measurements were performed from 5 to 340 K, in magnetic fields up to 5 Tesla, in a commercial SQUID magnetometer. The temperature dependence of the magnetization was measured following zero-field-cooling (ZFC) and field-cooling (FC) protocols (applied field H= 100 Oe). AC susceptibility measurements were performed at frequencies from 0.1 to $10^3$ Hz and a field amplitude of 3 Oe. SPA was measured using a commercial applicator (nB nanoscale Biomagnetics) working at 260 kHz and field amplitudes from 0 to 160 mT.

## 3. Experimental Results and Discussion.

Size histograms obtained from TEM images fit to a log-normal distribution. As an example, TEM images and size distributions of samples M03 and S01 are shown in figure 1a. The resulting $d_{TEM}$ and polidispersity, σ, of the three synthesized samples are shown in Table . As seen in figure 1, thermal decomposition of iron–organic precursors resulted in very uniform nanoparticles. Particles obtained by reflux in phenyl ether (samples M01 and M03) have a mean particle size of 5-7 nm and are less spherical than particles of sample S01, synthesized in 1-octadecene, which have an average particle size $d_{TEM}$ = 12.7 nm. The XRD pattern of the as prepared powder of sample S01 was indexed with the cubic spinel structure (figure 1b). In agreement with TEM data, the average crystallite size extracted from the most intense (3 1 1) reflection by applying the Scherrer formula was $d_{XRD}$ = 12 nm.

**Table 2:** Average particle diameter $d_{TEM}$, saturation magnetization $M_S$, coercive field $H_C$, Effective magnetic anisotropy $K_{eff}$ and Specific Power Absorption (SPA) of CoFe$_2$O$_4$ nanoparticles.

| Sample | $d_{TEM}$ (nm) | $T_B$ (K) | $H_C$ (Oe) 5 K | $H_C$ (Oe) 280 K | $M_S$ (emu/g) 5 K | $M_S$ (emu/g) 280 K | $K_{eff}$ (erg/cm$^3$) | SPA (W/g) |
|---|---|---|---|---|---|---|---|---|
| M01 | 5.7(3) | 174 | 14000 | 9.3 | 77.05 | 62.6 | $19 \times 10^6$ | 10.08 |
| M03 | 7.0(4) | 212 | 17100 | 8.3 | 73.19 | 60.0 | $14 \times 10^6$ | 18.59 |
| S01 | 12.7(4) | 315 | 17000 | 264 | 79.26 | 70.11 | $2.5 \times 10^6$ | 98.51 |

ZFC and FC curves are shown in figure 2a. In spite of the small size of the MNPs, large blocking temperatures $T_B$ were derived from the ZFC results, reflecting the large magnetocrystalline anisotropy of the particles. $T_B$ shifts to higher values as the particle size increases (see table 2). In the case of the S03 sample, an abrupt increase of the magnetization below $T_B$ was observed, which has not a clear explanation so far. In agreement with the high $T_B$ values, hysteresis loops measured at 5 K (see figure 2b) display high coercive fields, $H_C$, listed in table II. Above $T_B$, $H_C$ drops to zero as expected for the superparamagnetic regime. The values of saturation magnetization $M_S$ (see table II) are close to those of bulk $CoFe_2O_4$ [5].

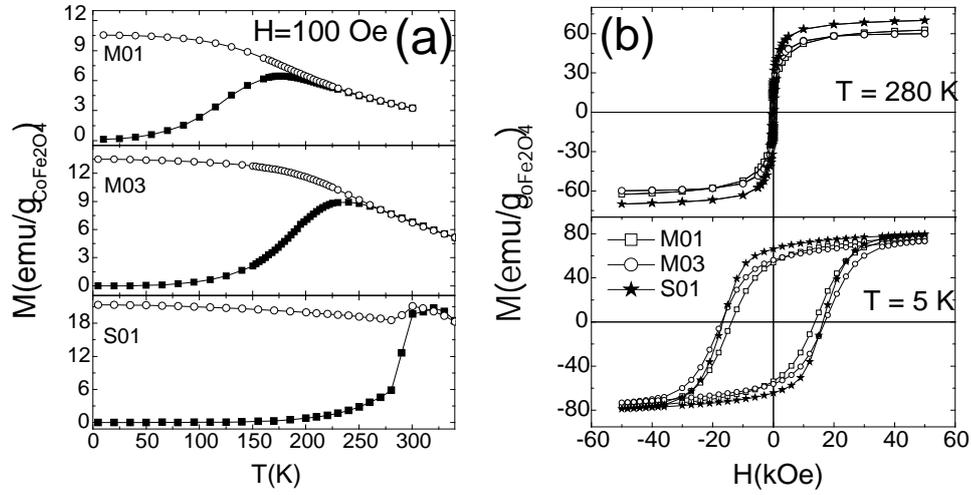

**Figure 2**. (a) M(T) data taken in Zero-field-cooled (ZFC) and field-cooled (FC) modes for $CoFe_2O_4$ samples. (b) M(H) curves measured at 280 K (top) and 5K (bottom).

In a single-domain particle with volume V and effective magnetic anisotropy $K_{eff}$, the reversion of the magnetic moment over the anisotropy energy barrier $E_a = K_{eff} V$ is assisted by thermal phonons, and the relaxation time $\tau$ is given by a Neél–Arrhenius law $\tau = \tau_0 \exp(E_a / k_B T)$. AC susceptibility results (figure 3a) showed that $\chi'(T)$ and $\chi''(T)$ exhibit the expected behaviour from SPM systems, i.e., the occurrence of a maximum at $T_m$, related to the unblocking of the magnetic moments. For the imaginary component $\chi''(T)$, the peak at $T_m$ corresponds to the temperature for which $f_{exp} = \tau$, where $f_{exp}$ is the ac excitation frequency. Therefore as the excitation frequency increases, $T_m$ shifts to higher values [6].

The linear dependence of $\ln(\tau)$ vs. $T_B^{-1}$ observed in figure 3b indicates that the Neél–Arrhenius model correctly suits the behaviour of the three samples. The fitting of the experimental data using the average particle diameter $d_{TEM}$ yielded the values of $K_{eff}$ listed in Table 2. The values for the smaller particles M01 and M03 are one order of magnitude larger than the first-order magnetocrystalline anisotropy constant of bulk magnetite ($K_1 = 2.0 \times 10^6$ erg/cm$^3$) [5], as found in other $CoFe_2O_4$ nanoparticles [7]. This anisotropy enhancement observed in MNPs is customarily associated to surface effects [8]. Accordingly, the larger particles S01 have $K_{eff}$ values close to the bulk value.

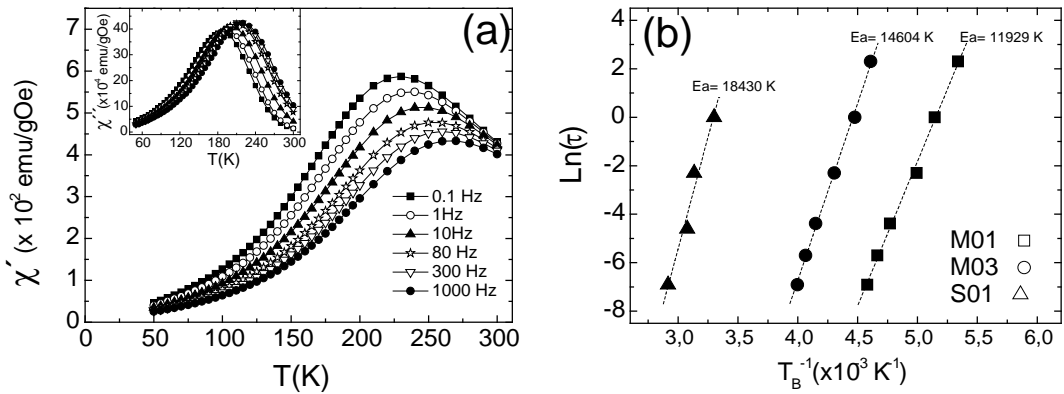

**Figure 3.** (a) Temperature dependence of the in-phase (real) component χ´(T) of the magnetic ac susceptibility for the M01 sample, at different excitation frequencies. Inset: Out of phase (imaginary) component χ´´(T). (b) Arrhenius plot of the relaxation time τ logarithm *vs.* the inverse blocking temperature $T_B^{-1}$.

To assess the influence of the size and magnetic parameters on heat generation, the SPA was measured from the temperature increase (ΔT) of a given mass of MNPs diluted in the liquid carrier during the time interval (Δt) of the experiment. The expression for power absorption Π per unit mass of the magnetic material is given by $\Pi = \frac{c_{LIQ}\delta_{LIQ}}{\phi}\left(\frac{\Delta T}{\Delta t}\right)$, where $c_{LIQ}$ and $\delta_{LIQ}$ are the specific heat capacity and density of the liquid carrier, respectively, and $\phi$ is the weight concentration of the MNPs in the colloid.[2] The SPA values listed in table 2 (in W/g of $CoFe_2O_4$) show that particles with $d_{TEM}$ = 12 nm (S01) are much more efficient for power absorption, whereas particles with 5-6 nm show poor heating aptitude. This results support the Néel relaxation-based mechanism of magnetic relaxation, in which the optimum size of nanoparticles for hyperthermia lays within a narrow size range, whose mean value strongly depends on the magnetic anisotropy of the system. [9] The large anisotropy of our NPs implies that good absorption can be obtained using smaller particles than for $Fe_3O_4$ particles, thus helping to the colloidal stability and size limitations for biomedical applications.

## 4. Conclusions

We have succeeded in producing highly stable, nearly monodisperse magnetic NPs with large effective magnetic anisotropy. Due to this anisotropy, high blocking temperatures were observed even in small (5-6 nm) particles. Moreover, excellent heating efficiency was observed for NPs with d =12-13 nm, much smaller than the optimal values needed to obtain similar SPA values using magnetite NPs (25-35 nm). Our results suggest that $CoFe_2O_4$ NPs could overcome some of the problems derived from colloidal instability of large $Fe_3O_4$ nanoparticles for biomedical applications.

## 5. Acknowledgements

The authors acknowledge financial support from the Spanish Ministry of Science and Innovation (MICINN) (projects CONSOLIDER CSD2006-00012, MAT2008-0274/NAN, MAT2008-06567-C02-02/NAN) and Diputación General de Aragón (project PI118/08). GFG thanks MICINN through the Ramon y Cajal program. AGR thanks CIBER-BBN center of the Spanish Ministry of Health through projects IMAFEN, NANOMAG, MICROPLEX and MONIT.